\documentclass{PoS}

\usepackage{epstopdf}
\usepackage{aas_macros}
\usepackage{subfig}

\def\gs{\Upsilon_\star}

\title{Dark Matter Phase Transition Constrained at $\mathcal{O}(0.1)$ eV with LSB Rotation Curves}

\ShortTitle{DM Phase Transition Constrained at O(0.1) eV with LSB Rotation Curves}

\author{\speaker{JORGE MASTACHE}%
         \thanks{I would like to thank the organizers and staff for giving me the opportunity to present my work in this excellent congress.}\\
        Instituto de Fisica, Universidad Nacional Autonoma de Mexico, Apdo. Postal 20-364, 01000, Mexico, D.F.\\
        E-mail: \email{mastache@fisica.unam.mx}}
\author{Axel de la Macorra\\
        Instituto de Fisica, Universidad Nacional Autonoma de Mexico, Apdo. Postal 20-364, 01000, Mexico, D.F.\\
        E-mail: \email{macorra@fisica.unam.mx}}
\author{Jorge L. Cervantes-Cota\\
        Depto. de Fisica, Instituto Nacional de Investigaciones Nucleares, Col. Escandon, Apdo. Postal 18-1027, 11801, D.F. Mexico.\\
        E-mail: \email{jorge.cervantes@inin.gob.mx}}

\abstract{In order to unravel the nature of the dark matter (DM) we have proposed a particle-physics motivated model called Bound Dark Matter (BDM) that consist in DM massless particles above a threshold energy ($E_c$) that acquire mass below it due to nonperturbative methods. Therefore, the BDM model describes DM particles which are relativistic, hot dark matter (HDM) in the denser (inner) regions of galaxies and describes nonrelativistic, cold dark matter (CDM) where halo density is below $\rho_c\equiv E_c^4$. We test this model by fitting rotation curves from Low Surface Brightness (LSB) galaxies from The HI Nearby Galaxy Survey (THINGS). We use a particular DM cored profile that contains three parameters: a typical scale length ($r_s$) and density ($\rho_0$) of the halo, and a core radius ($r_c$) stemming from the relativistic nature of the BDM model. Since the energy $E_c$ parameterizes the phase transition due to the underlying particle physics model, it is independent on the details of galaxy and/or structure formation and therefore the DM profile parameters $r_s, r_c, E_c$ are constrained, leaving only two free parameters. Through the results we agree with previous ones implying that cored profiles are preferred over the N-body motivated cuspy profiles. We also compute 2D likelihoods of the BDM parameters $r_c$ and $E_c$ for the different galaxies and matter contents, and find an average galaxy core radius $r_c = 1.48$ kpc and a transition energy between hot and cold dark matter at $E_c = 0.06^{+0.07}_{-0.03} {\rm \ eV}$. The phase transition scale $E_c$  is a new fundamental scale for our DM model well motivated theoretical origin from gauge group dynamics.
}

\FullConference{VIII International Workshop on the Dark Side of the Universe\\
                 June 10-15, 2012\\
                 Búzios, Rio de Janeiro, Brasil}

\begin{document}

\section{Introduction}\label{sec:introduction}
It has been known for several decades that the kinematics of disc galaxies exhibit a mass discrepancy \cite{1978PhDT195B,BosmaKruit,Rubin:1980zd}. The distribution of the baryonic components cannot justify the measured circular velocities of the rotation curves (RC)\cite{Gentile:2006hv,Gentile:2007sb}. This is usually solved by adding an extra mass component, the dark matter (DM) halo. Understanding the amount and distribution of DM in galaxies has been a major work in several years and more recently \cite{Gentile:2006hv, McGaugh:2001yc, deBlok:2002tg, Gentile:2004tb, Gentile:2005tu,deBlok:2008wp}. Mass models are fitted to galaxy rotation curves in order to study the systematic properties of halos specially in (dwarf and early type) Low Surface Brightness (LSB) galaxies, because visible matter contributes only a small fraction of the total mass, therefore it is believed they are dominated by DM.

The halo parameters are derived by fitting particular profiles. There are essentially two types of profiles, the ones emerging in cosmological $N$-body simulations of structure formation in the framework of the $\Lambda CDM$ theory \cite{Navarro:1995iw, Moore:1999gc, Neto:2007vq, Fukushige:2000ar}. that favour cuspy halo profiles, peaked in its inner region. On the other hand, the phenomenological motived cored profiles that challenge the CDM theory furthermore it is supported by many observations, such as the Burkert or Pseudo-Isothermal (ISO) profiles \cite{Burkert:1995yz}.

Cuspy and cored profiles can both be fitted to most LSB rotation curves. The cuspy profiles that do fit to galaxies in many cases give systematically unphysical concentrations ($c$), too high velocities $V_{200}$ and/or low values for the stellar mass-to-light ratio, as result of its steep rise in the center. On the other hand the cored DM halo provides a very satisfactory fit to the observed data, in particular to the central part for such slowly rising rotation curves and, more generally, superior to that obtained by assuming a NFW profile. It is an aim of this paper to directly test the NFW halo profile, and compare its fits with an alternative framework of the cored halo profiles.

In addition, the recent analysis of the CMB data  with measurements of BAO and the Hubble constant reported an excess of the effective number of relativistic degrees of freedom, $N_{eff} = 4.6 \pm 0.8$, in consistency with Ref. \cite{Komatsu:2010fb}, opening the possibility to have extra nonstandard-model relativistic degrees of freedom, as our BDM particle.

We organized this presentation as follows: In Sec. \ref{sec:PartModel} we explain the particle model and present our BDM profile, which has NFW as a limit at low energies. In Sec. \ref{sec:MassModel} we describe the galaxy sample considered for the rotation curves study. The different mass models and components (gas, stars, DM halo) are presented. The results and conclusions are presented in Sec. \ref{sec:results}.

\section{Particle Model}\label{sec:PartModel}

Cosmological evolution of gauge groups, similar to QCD,  have been studied to understand the nature of dark energy and DM \cite{delaMacorra:2001ay,delaMacorra:2002tk}. In order to determine the nature of the dark matter particle one of us recently proposed a particle physics model called bound dark matter (BDM) \cite{delaMacorra:2009yb}. In this model the fundamental particles have a vanishing masses that acquired mass due to nonperturbative gauge mechanism at a threshold scale $E_c$. The mass of this particles is expected to be at the same order of magnitude as the phase transition scale such that $m_{BS}=d\,E_c$ with $d=O(1)$ being a proportionality constant.

There are two natural places where one may encounter the dark matter transition. One is at early cosmological times and the second place is at the inner regions of galaxies. In the last case, the BDM particles at high densities, inner regions, are relativistic, i.e. Hot DM (HDM), but in the outer regions they behave as standard CDM. To realize this idea we match the theoretically mass model with observation using a particular DM cored profile (Eq. \ref{eq:rho_bdm}) that contains three parameters: a typical scale length ($r_s$) and density ($\rho_0$) of the halo and a core radius ($r_c$) stemming from the relativistic nature of the BDM model. The last parameter $r_c$ help us to determine if the galaxies prefer cored or cuspy halos such that when $r_c \ne 0$ implies the existence of a constant central surface density of DM in galaxies and does not depend on which specific cored density profile is assumed. The galactic core density is given by $\rho_c \equiv E_c^4 = \rho_0 r_s / 2 r_c$ and the profile properties will determine the energy scale of the particle physics model.

\begin{equation}\label{eq:rho_bdm}
\rho_{bdm} = \frac{2 \rho_c}{\left( 1+\frac{r}{r_c}\right)\left( 1+\frac{r}{r_s}\right)^2},
\end{equation}

The $E_c$ parameter is a new fundamental scale for the DM which can be theoretically determined using gauge group dynamics, i.e. it is independent of the properties of galaxies, e.g. their luminosity, once the gauge group is known. Our goal here is to analyze the LSB galaxies nevertheless we expect the same results from high-surface brightness (HSB) galaxies. It has been pointed out \cite{Persic:1995ru, Disney:2008zz, 2008Natur.455.1049V} that basic galaxy parameters as mass, size, baryon-fraction, central density, are not independent from each other but in fact all of them do depend on one parameter that works as a galaxy identifier, it will be interesting if the relation between $E_c$ and other scale-relations could be found in future works.

However, even though we propose $E_c$  as a new fundamental constant for DM, close related to the mass of our DM particle, and  it is important to note that its value is not yet known and we require observational evidence to determine  it.  The same is true for all masses of the standard model of particles, i.e. their value is not predicted by the standard model and it is the experimental results that fixed them in a consistent manner. We use here the information on galactic rotation curves to determine $E_c$. However, the extracting of $E_c$  does depend not only on the quality of the observational data, the mass models used but also on the choice of DM profile.

\section{Mass model}\label{sec:MassModel}
To theoretical fit the observational data, we use The HI Nearby Galaxy Survey (THINGS), which collects high resolution $HI$, $H_{\alpha}$ and optical data far enough of the rotation curve, that all the free parameters can be measured \cite{Walter:2008wy} and can help to distinguish among the different DM profiles. Given these properties it is adequate to test the above-mentioned DM profiles with THINGS, which has been used to test different core/cusp profiles. In the cases where some of the galaxies have already been modelled by other authors, the results tend to agree pretty well.

We studied the rotation curves using BDM, NFW, Burkert and ISO DM profiles, taking into account the contribution from different mass models: i) DM alone; ii) DM and HI disc; and iii) DM, HI disc and the stellar disk.

The kinematics of stars bring a very challenging problem in the analysis, mainly due to the uncertainty of the mass-to-light ratio ($\gs$) and to its dominant behavior close to the galactic center. Some considerations have been made in order to reduce this uncertainty in the parameters \cite{Salpeter:1955it, Kroupa:2000iv}, but still the stellar contribution is not well known. In our case we studied three different $\gs$ models. In addition to the standard method in which the stellar $\gs$ is model-independent, we also modelled their RCs by assuming for the latter quantity the values obtained from the galaxy colours as predicted by spectro-photometric models with a (i) diet-Salpeter or a (ii) Kroupa initial mass function (IMF). The resulting mass models well reproduce the RCs, and the relative halo parameters are derived within a reasonable uncertainty

When considering a Kroupa or diet-Salpeter mass model we compute the rotation curve of the stars from the brightness distribution taking into account that the $\gs$ is function of the radius in order to consider the different stars contribution as it depends on the region that we were analyzing. We have not include the contribution of the molecular gas since its surface density is only a few percent of the that of the stars, therefore its contribution is reflected in a small increase in $\gs$

For each of the 17 galaxies studied (D 154, I 2574, N 925, N 2366, N 2403, N 2841, N 2903, N 2976, N 3031, N 3198, N 3521, N 3621, N 4736, N 5055, N 6946, N 7331, and N 7793), we have applied the best-fitting model technique to all of the profiles. It consists in minimizing the $\chi^2$ in the three dimensions of the parameter space defined by ($\rho_0$, $r_0$, $r_s$). Whenever a model independent is considered (Free $\gs$) a fourth dimension is added, and when the bulge component is taken into account, a fifth dimension parameter space is considered. The method explain here in better described in \cite{Mastache:2011cn}.

The most inner part of the data enable us to compute the BDM parameters $r_c$ and $E_c$, both related to the galactic core. We present the 1$\sigma$ and $2\sigma$ likelihood  contour plots of these parameters for the different galaxies and the most realistic (Kroupa) and the upper limit (only DM) mass model. We notice that being $r_c$ different for each mass model it is consistent within the $2\sigma$ error for each galaxy.

\section{Results}\label{sec:results}
\begin{figure}
  \centering
    \subfloat[\footnotesize{$R_c vs E_c$ : Minimal Disk}]{ \includegraphics[width=0.50\textwidth]{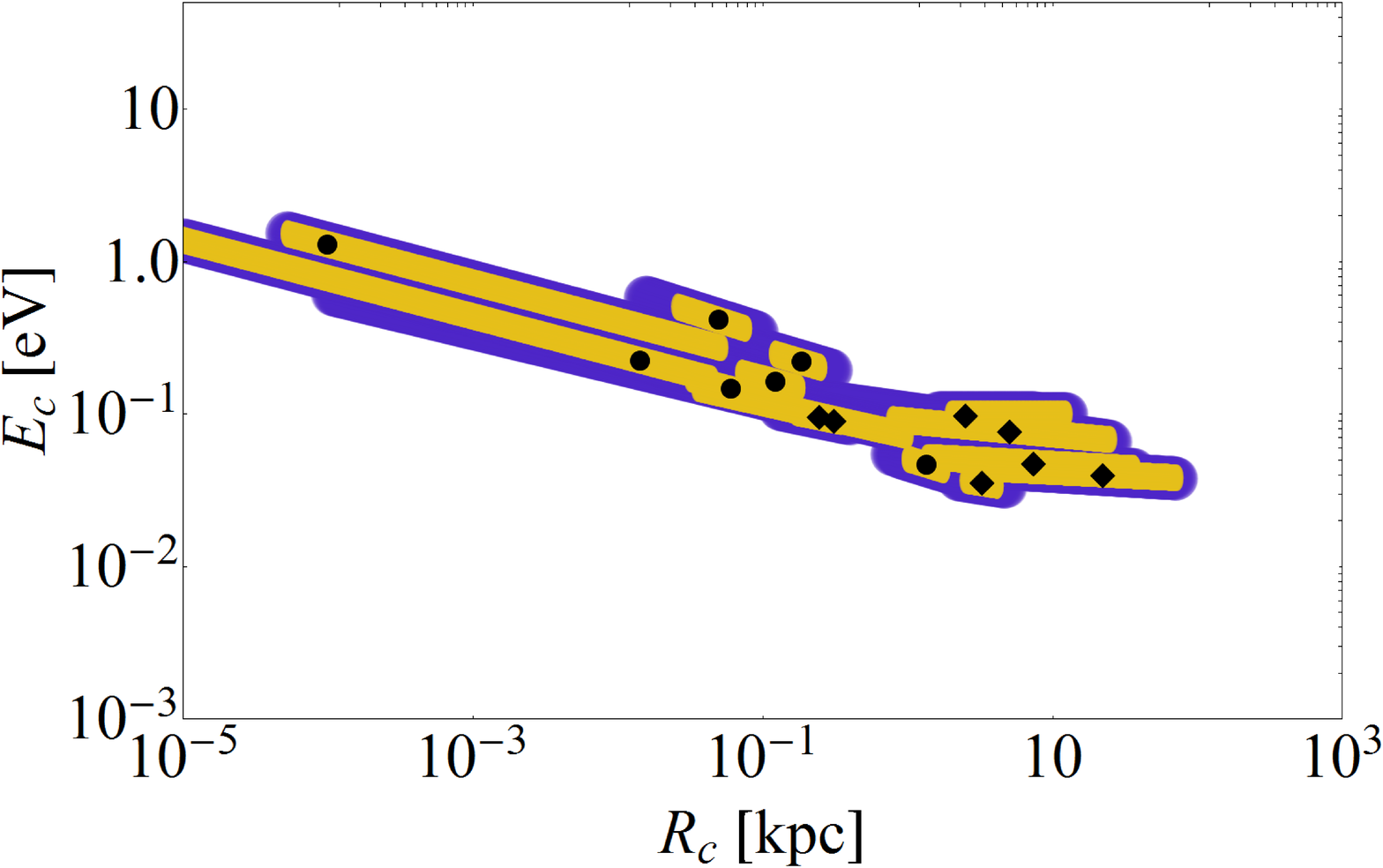}}
    \subfloat[\footnotesize{$r_c$ vs $E_c$ : Kroupa}]{ \includegraphics[width=0.50\textwidth]{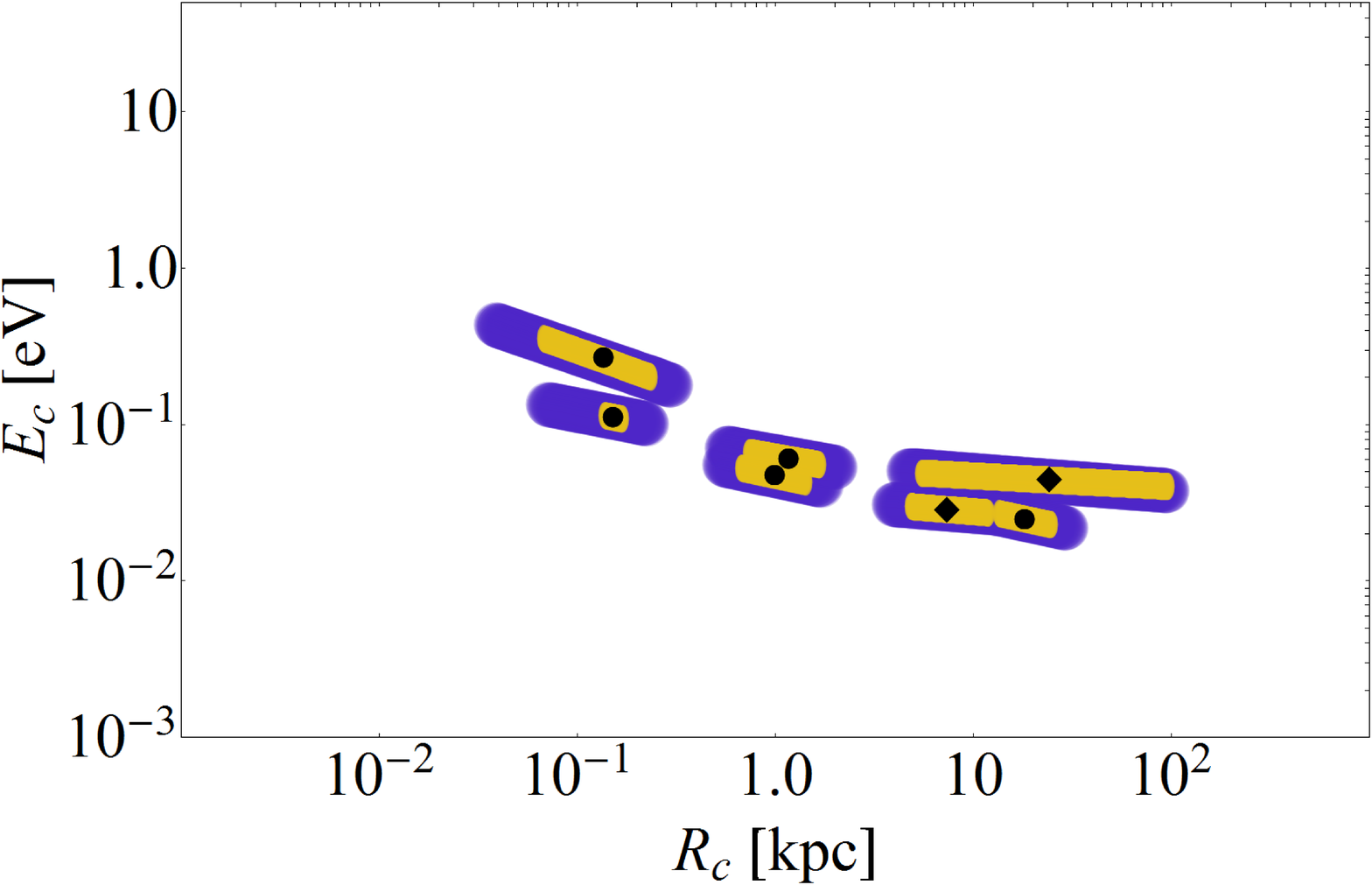}}
    \caption{\footnotesize{
        We plot the values of $r_c$ vs $E_c$ for each galaxy and different mass models. Circles represent galaxies with $r_c\ll r_s$, diamond markers are galaxies corresponding $r_c = r_s$. In all cases the areas in yellow and blue represent the $1\sigma$ and $2\sigma$ confidence levels. }}
    \label{fig:RcvsEc}
\end{figure}
The main result of this study is that DM satisfies the well defined energy scale $E_c$.  We obtain the upper limit from the LSB rotation curves that the phase transition energy scale $E_c$, between HDM and CDM for our BDM profile, is at $E_c = 0.11^{+0.21}_{-0.07} {\rm \ eV}$ and a core radius is the order of $r_c \sim 300 {\rm \ pc}$ when we consider the halo dark matter as the only mass component, see Fig. \ref{fig:RcvsEc}. On the other hand, we found an energy $E_c = 0.06^{+0.07}_{-0.03}$ and a core $r_c = 1.48$ when considering gas, DM halo, and the minimum contribution of stars (Kroupa). The observed kinematics of Local Group dwarf spheroidal galaxies are consistent with this value but should be confirmed with a larger sample. Moreover, the BDM profile fits equally well or better than other cored profiles and only for a few LSB galaxies our model resembles that of NFW.

It is also interesting to note the coincidence in the magnitude of the sum of neutrino masses with the magnitude of $E_c$ obtained in this analysis, and how this  could  open an interesting  connection between the generation of DM and neutrinos masses, but this will be presented elsewhere \cite{NBDM}. The magnitude of  the energy transition $E_c$  is of the same order as the upper limit of the total mass of the neutrinos $\sum m_\nu < 0.58 {\rm \ eV} (95\% {\rm \ CL)}$ from the seven year Wilkinson Microwave Anisotropic Probe (WMAP), Baryon Acoustic Oscillation (BAO), and the Hubble constant ($H_0$) \cite{Larson:2010gs, Komatsu:2010fb}.

A value of $E_c$  can be considered as a new scaling law, which remind us the correlation found by other authors [\cite{Kormendy:2004se, Donato:2009ab}and references there in] that the product of the core radius and the central density of the halo (for a pseudo-isothermal profile) is nearly constant despise of their luminosity and Hubble type of the galaxy. In which case we extend this conclusion by theoretical explaining the reason why this product remain constant through different galaxies. Nevertheless this remain to be explored. We also confirms a the result already claimed by other authors about the shape of the DM halo in the center of LSB galaxies, namely that its density profile is probably closer to an cored profile than to a cuspy profile.

\bibliographystyle{unsrt}
\bibliography{skeleton}


\end{document}